\documentclass[%
reprint,
letterpaper,
nobibnotes,
bibnotes,
 amsmath,amssymb,
 aps,
 nobalancelastpage,
]{revtex4-1}

\usepackage[pdftex]{graphicx}
\usepackage{siunitx}
\usepackage[usenames]{xcolor}
\usepackage{mathtools}

\usepackage[normalem]{ulem}
\usepackage{MnSymbol}
\usepackage{wasysym}

\makeatletter
\newcommand\footnoteref[1]{\protected@xdef\@thefnmark{\ref{#1}}\@footnotemark}
\makeatother

\usepackage{amssymb,amsfonts,amsmath}

\usepackage{calc}

\newcommand{\gdot}{\dot{\gamma}}
\newcommand{\Abar}{$\bar{\rm A}$}
\newcommand{\Bbar}{$\bar{\rm B}$}
\newcommand{\fAbar}{f_{\bar{\rm A}}}
\newcommand{\fBbar}{f_{\bar{\rm B}}}
\newcommand{\phiA}{\phi_{\rm A}}
\newcommand{\phiAbar}{\phi_{\bar{\rm A}}}

\newcommand{\phiAB}{\phi_{\rm AB}}
\newcommand{\phiAbarB}{\phi_{\bar{\rm A} {\rm B}}}
\newcommand{\phiABbar}{\phi_{{\rm A} \bar{\rm B}}}
\newcommand{\phiAbarBbar}{\phi_{\bar{\rm A} \bar{\rm B}}}
\begin{document}

%\preprint{APS/123-QED}

\title{A constraint-based approach to granular dispersion rheology}
\author{B. M. Guy}
\email{b.m.guy1990@gmail.com}
\author{J. A. Richards}
\email{jamesrichards92@gmail.com}
\author{D. J. M. Hodgson}
\author{E. Blanco}%
\author{W. C. K. Poon}
\affiliation{SUPA, School of Physics and Astronomy, The University of Edinburgh, King's Buildings, Peter Guthrie Tait Road, Edinburgh, EH9 3FD, United Kingdom}

\date{\today}

\begin{abstract}
We present a phenomenological model for granular suspension rheology in which particle interactions enter as constraints to relative particle motion. By considering constraints that are \emph{formed} and \emph{released} by stress respectively, we derive a range of experimental flow curves in a single treatment and predict singularities in viscosity and yield stress consistent with literature data. Fundamentally, we offer a generic description of suspension flow that is independent of bespoke microphysics.
\end{abstract}

\maketitle

Concentrated particulate dispersions are ubiquitous in industry. When the particle size is in the granular (i.e., non-Brownian) regime (radius $R \gtrsim \SI{1}{\micro\metre}$), their flow is notoriously difficult to predict and control \cite{mewis2012colloidal, denn2014rheology}.
Paradoxically, a suspension of non-Brownian hard particles has no intrinsic time or stress scale and so should have a viscosity $\eta$ that is independent of shear stress $\sigma$ and rate $\gdot$ \cite{denn2014rheology,cates2014granulation}. In reality, three classes of flow curve $\eta(\sigma)$ are observed, none of which is Newtonian. Some granular suspensions shear thin ($d\eta/d\sigma<0$, class 1) \cite{zhou1995yield,heymann2002solid}, others shear thicken ($d\eta/d\sigma>0$, class 2) \cite{Barnes1989,guy2015towards,royer2016rheological} while others show a varied combination of thinning and thickening (class 3): thinning then thickening (class 3a) \cite{brown2014shear,brown2010generality}, thickening then thinning (class 3b) \cite{gamonpilas2016shear,zarraga2000characterization,dai2013viscometric} or more complex behavior \cite{brown2010generality,bertrand2002shear,chatte2018shear} (class 3c). In each class, the suspensions can become solid-like \cite{peters2016direct} or flow unstably \cite{fall2015macroscopic,hermes2016unsteady}.

Such behavior likely stems from details of the particle interactions \cite{denn2014rheology} set by, e.g., surface chemistry \cite{galvez2017dramatic} or roughness \cite{hsiao2017rheological}. Most models incorporate such interactions in a bespoke manner. Notably, a phenomenological model by Wyart and Cates (WC) \cite{wyart2014discontinuous} predicts thickening (class 2) due to a transition from frictionless (static friction coefficient $\mu \approx 0$) to frictional ($\mu>0$) particle contacts above a critical ``onset stress". Atomic force microscopy confirms this picture for several systems \cite{comtet2017pairwise,chatte2018shear} and the WC model fits a number of experimental flow curves \cite{guy2015towards,royer2016rheological,hermes2016unsteady}; although, quantitative discrepancies with microscopic simulations remain \cite{denn2018shear}.
 
To recast the WC model within a more general framework, recall that frictional contacts constrain inter-particle sliding. Crucially, the WC model is agnostic to the exact mechanism by which sliding is constrained, so that disparate microphysics, e.g., stress-induced interlocking of asperities ~\cite{hsiao2017rheological,hsu2018roughness}, hydrogen bonding \cite{james2017interparticle} or `traditional' Coulomb friction can all give rise to the same macroscopic, shear-thickening phenomenology.

In this broader framework, the WC model deals with a single type of constraint: sliding. Rolling (rotations about axes perpendicular to the line of centres) and twisting (rotations about the line of centres) degrees of freedom remain unconstrained. By assuming that sliding constraints are formed at increasing stress, the WC model accounts for class 2 behavior, which, however, is rare in practice. %Most, if not all, literature examples are specially synthesised model systems. 
Real systems are typically class 1 or 3, for which current explanations involve the {\it ad hoc} ``bolting together" of different kinds of bespoke physics \cite{brown2010generality}.

Here, we generalize the WC model to \emph{two} constraint types, $\mathcal{A}$ and $\mathcal{B}\neq\mathcal{A}$. For example, $\mathcal{A}={\rm sliding}$ and $\mathcal{B}={\rm rolling}$ (e.g., due to adhesive contact \cite{dominik1995resistance}). Constraint $\mathcal{A}$ is \emph{formed} by stress, while constraint $\mathcal{B}$ is \emph{released} by stress. Now, one single model predicts all observed classes of experimental flow curves. This includes class 3b, whose rheology we find is sensitive to the exact interplay of constraint formation and release, thus explaining the variability of such systems observed experimentally. Moreover, we make non-trivial predictions for the emergence of singularities that are consistent with literature data, e.g., re-entrant jamming in class 3a and a yield stress diverging below random close packing for class 1.

%------------------------------------------------THEORY -- TYPE-A only-------------------------------------------------%

We begin by considering the case of a single constraint type, $\mathcal{A}$, which is \emph{formed} by increasing stress $\sigma$. In the original WC model, $\mathcal{A}={\rm sliding}$. There are two possible contact states: one in which $\mathcal{A}$ is constrained, \Abar, and one in which $\mathcal{A}$ is unconstrained, A. (Thus, e.g., \Abar$=$`non-sliding' and A$=$`sliding'.) These states are associated with different jamming points, $\phiA$ and $\phiAbar$; thus, e.g., a suspension with all contacts in state A jams ($\eta \rightarrow \infty$) at $\phiA$.
Existing literature suggests that $\phiAbar <\phiA$ \cite{mari2014shear,henkes2010critical,liu2017equation}. Thus, for monodisperse spheres with $\mathcal{A}={\rm sliding}$, $\phiA\approx \phi_{\rm rcp}\approx 0.64$ (random close packing) and $\phiAbar \approx \phi_{\rm rlp}\approx 0.55<\phiA$ \cite{mari2014shear} (random loose packing \cite{silbert2010jamming}). 

Under shear, $\mathcal{A}$ is constrained in a $\sigma$-dependent fashion, so that a fraction $\fAbar$ of the contacts are in the \Abar\, state, with $d\fAbar/d\sigma>0$. A suspension with $\fAbar=0$ (all contacts in state A) jams at $\phiA$; when $\fAbar=1$ (all contacts in state \Abar), it jams at $\phiAbar$. For $0<\fAbar<1$, the system jams at an intermediate volume fraction, $\phi_{\rm J}(\fAbar)$, the functional form of which is not known for all constraint types. We use WC's form for constrained sliding \cite{wyart2014discontinuous}:
\begin{equation}
    \phi_{\rm J}(\fAbar)=\fAbar \phiAbar + (1-\fAbar)\phiA.
    \label{eq:phiJ1}
\end{equation}

The viscosity $\eta$ increases as the distance to jamming, $\Delta \phi=\phi_{\rm J}(\fAbar)-\phi$, decreases, diverging as $\phi \rightarrow \phi_{\rm J}$.  Again, we use WC's form for sliding constraints \cite{wyart2014discontinuous}:
\begin{equation}
    \eta(\phi,\phi_{\rm J})=\eta_0\phi_{\rm J}^2(\Delta \phi)^{-2}=\eta_0 [1-\phi/{\phi_{\rm J}}]^{-2},
    \label{eq:eta}
\end{equation}
where $\eta_0$ is the viscosity of the suspending medium.

This single-constraint scenario leads to shear thickening: increasing $\sigma$ increases $\fAbar$, decreasing $\phi_{\rm J}$ (and $\Delta \phi$), Eq.~(\ref{eq:phiJ1}), driving up $\eta$, Eq.~(\ref{eq:eta}). The exact form of the flow curve $\eta(\sigma)$ depends on $\fAbar(\sigma)$, which encapsulates the stress-dependent microphysics between particles.

%------------------------------------------------THEORY -- TYPE-A and TYPE-B -------------------------------------------------%

Now introduce a second constraint type $\mathcal{B}\neq\mathcal{A}$ that is \emph{released} by stress, giving four contact states: AB, \Abar B, A\Bbar~and \Abar\Bbar~(where \Bbar/B means $\mathcal{B}$~is/is not constrained), each with an associated jamming point: $\phiAB$, $\phiAbarB$, $\phiABbar$~ and $\phiAbarBbar$. Random close packing at $\phiAB = \phi_{\rm rcp}$ is the same for all identities of $\mathcal{A}$ and $\mathcal{B}$, since both are unconstrained in the AB state.
The other jamming points depend on the nature of $\mathcal{A}$ and $\mathcal{B}$, and are unknown in general. In static packings of dry grains, combining multiple constraint types typically lowers the minimum packing fraction for mechanical stability \cite{gilabert2007computer, estrada2008shear, liu2017equation}. We suppose that this also applies to the jamming point of sheared suspensions, so that $\phiAbarBbar < \phiAbarB, \phiABbar < \phiAB$. 

In a fraction $\fBbar$ of the contacts, $\mathcal{B}$ is constrained. Importantly, $d\fBbar/d\sigma<0$. The jamming volume fraction, $\phi_{\rm J}$, now depends on both $\fAbar$ and $\fBbar$, with $\phi_{\rm J} \rightarrow \phiAB$ at $(\fAbar,\fBbar)=(0,0)$, $\rightarrow \phiAbarB$ at $(\fAbar,\fBbar)=(1,0)$, $\rightarrow \phiABbar$ at $(\fAbar,\fBbar)=(0,1)$ and $\rightarrow \phiAbarBbar$ at $(\fAbar,\fBbar)=(1,1)$. The simplest functional form consistent with these limits is:
\begin{eqnarray}     
     \phi_{\rm J}(\fAbar,\fBbar) = (1-\fAbar)(1-\fBbar)\phiAB + \nonumber \\ \fAbar (1-\fBbar)\phiAbarB +  (1-\fAbar)\fBbar \phiABbar +  \fAbar \fBbar \phiAbarBbar.                                                                                                                                                   
     \label{eq:phiJ2}
\end{eqnarray}
Finally, we again relate $\phi_{\rm J}$ to $\eta$ via Eq.~(\ref{eq:eta}).

\begin{figure}[t]
\centering
\begin{minipage}{0.49\columnwidth}
\includegraphics[width=1.02\textwidth]{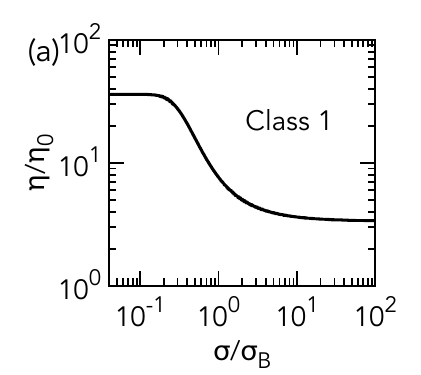}
\end{minipage}
\begin{minipage}{0.49\columnwidth}
\includegraphics[width=\textwidth]{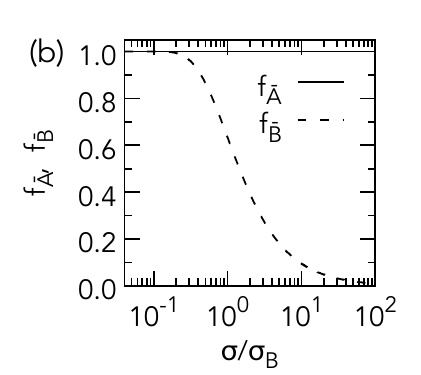}
\end{minipage}
\\
\begin{minipage}{0.49\columnwidth}
\includegraphics[width=1.02\textwidth]{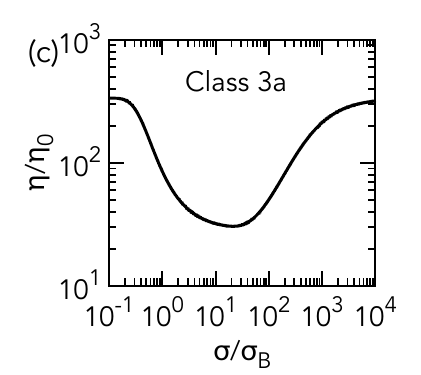}
\end{minipage}
\begin{minipage}{0.49\columnwidth}
\includegraphics[width=1.02\textwidth]{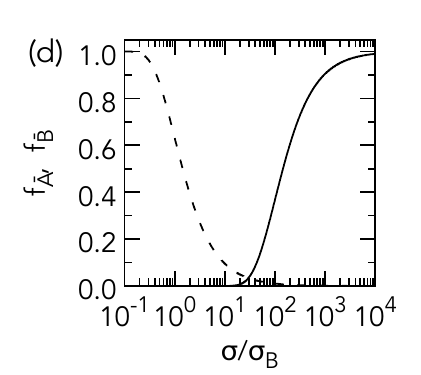}
\end{minipage}
\\
\begin{minipage}{0.49\columnwidth}
\includegraphics[width=1.02\textwidth]{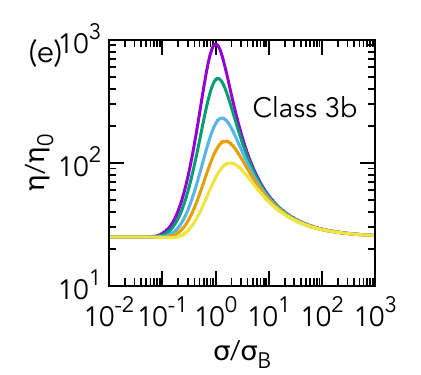}
\end{minipage}
\begin{minipage}{0.49\columnwidth}
\includegraphics[width=1.02\textwidth]{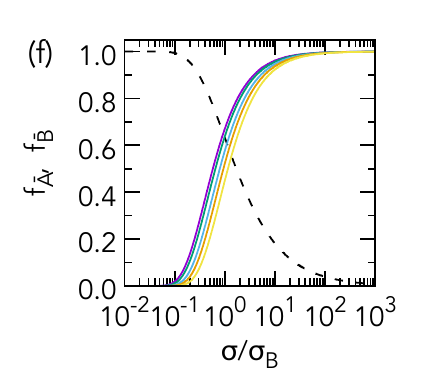}
\end{minipage}
\caption{Two-constraint model: example flow curves. In all panels, $\phiAbarB=\phiABbar=0.86\phiAB$ and $\phiAbarBbar=0.47\phiAB$. (a,c,e) $\eta/\eta_0$ versus $\sigma/\sigma_{\rm B}$. (b,d,f) $\fAbar$ (solid lines) and~$\fBbar$ (dashed lines) versus $\sigma/\sigma_{\rm B}$. Parameters for class 1 (a-b):  $\sigma_{\rm A}/\sigma_{\rm B}=10^{-6}$, $\alpha=\beta=1$ and $\phi=0.39\phiAB$; class 3a (c-d): $\sigma_{\rm A}/\sigma_{\rm B}=10^2$, $\alpha=\beta=1$ and $\phi=0.81\phiAB$; class 3b (e-f): $\alpha=1$, $\beta=0.7$, $\phi=0.69\phiAB$ at different $\sigma_{\rm A}/\sigma_{\rm B}=0.40$ (purple), 0.45 (green), 0.55 (cyan), 0.65 (orange) and 0.80 (yellow), from top to bottom in (e) and left to right in (f).}
\label{fig:GWC}
\end{figure}

The rheology for the two-constraint case is far richer than that for a single constraint. Even rather bland choices for $\fAbar(\sigma)$ and $\fBbar(\sigma)$ readily lead to flow curves of all classes, Fig.~\ref{fig:GWC}, including the little-understood class 3b. Within each class, particularly class 3, the exact phenomenology is sensitive to the detailed interplay between $\fAbar$ and $\fBbar$, i.e., to the relative formation of type-$\mathcal{A}$ and release of type-$\mathcal{B}$ constraints with stress.

We present predictions using for $\fAbar(\sigma)$:
\begin{equation}
    \fAbar(\sigma)=\exp[-(\sigma_{\rm A}/\sigma)^\alpha],
    \label{eq:fAbar}
\end{equation}
where $\sigma_{\rm A}$ is the characteristic stress for the \emph{formation} of type-$\mathcal{A}$ constraints and $\alpha$ controls the rapidity of type-$\mathcal{A}$ constraint formation with $\sigma$, and for $\fBbar(\sigma)$:
\begin{equation}
    \fBbar(\sigma)=1-\exp[-(\sigma_{\rm B}/\sigma)^{\beta}],
    \label{eq:fBbar}
\end{equation}
where $\sigma_{\rm B}$ is the characteristic stress for the \emph{release} of type-$\mathcal{B}$ constraints and $\beta$ controls the rapidity of type-$\mathcal{B}$ constraint release. The rheological class of the system is now controlled by $\sigma_{\rm A}/\sigma_{\rm B}$ and $\alpha/\beta$. The exact scenario depends on the relative values of the jamming points; however, the qualitative phenomenology is the same provided $\phiAbarBbar < \phiAbarB, \phiABbar < \phiAB$. To make concrete predictions, we use $\phiAbarB=\phiABbar=0.86\phiAB$ and $\phiAbarBbar=0.47\phiAB$ (so if $\phiAB=\phi_{\rm rcp}=0.64$, $\phiAbarB=\phiABbar=0.55$ and $\phiAbarBbar=0.30$).

Class 1 (shear thinning) arises whenever $\sigma_{\rm A}/\sigma_{\rm B} \ll 1$, independent of $\alpha/\beta$, Fig.~\ref{fig:GWC}(a-b) (see caption for parameters). Since $\sigma_{\rm A}$ is readily exceeded, $\mathcal{A}$ is always constrained and $\fAbar=1$, Fig.~\ref{fig:GWC}(b). The shape of the flow curve $\eta(\sigma)$, Fig.~\ref{fig:GWC}(a),  reflects $\fBbar(\sigma)$: it shear thins as type-$\mathcal{B}$ constraints are progressively released, reaching a Newtonian plateau when they are all removed.

Class 3a (thinning then thickening) arises when $\sigma_{\rm A}/\sigma_{\rm B} \gg 1$ and $\alpha/\beta \leq 1$, Fig.~\ref{fig:GWC}(c-d). $\sigma_{\rm A}$ and $\sigma_{\rm B}$ are well separated, Fig.~\ref{fig:GWC}(d), so type-$\mathcal{B}$ constraints are almost completely released before type-$\mathcal{A}$ ones begin to form. $\eta(\sigma)$ first shear thins as $\fBbar$ drops, then shear thickens as $\fAbar$ subsequently rises. If $\sigma_B=0$, i.e., $\mathcal{B}$ is always unconstrained, simple shear thickening (class 2) is recovered. 

Class 3b  (thickening then thinning) occurs only when $\sigma_A/\sigma_B\sim 1$, Fig.~\ref{fig:GWC}(e-f). Now, the form of $\eta(\sigma)$ cannot be easily deduced from the evolution of $\fAbar(\sigma)$ and $\fBbar(\sigma)$, and the viscosity ``peak" is profoundly sensitive to changes in $\sigma_{\rm A}/\sigma_{\rm B}$. Thus, for $\alpha/\beta=1.4$ and $\phi/\phiAB=0.69$, halving $\sigma_{\rm A}/\sigma_{\rm B}$ drops the peak by a factor of ten. There is a similar sensitivity to $\alpha/\beta$ at fixed $\sigma_{\rm A}/\sigma_{\rm B}$, which we explore in the Supplementary Information (SI) \cite{SI}. In each case, however, the extended region of shear thinning after the peak is controlled by the progressive release of constraints on $\mathcal{B}$, $\fBbar \rightarrow 0$, with constraints on $\mathcal{A}$ almost saturated, $\fAbar \approx 1$.

Such sensitivity is consistent with experiments, which find class 3b $\eta(\sigma)$ sensitive to small changes in, e.g., particle size \cite{gamonpilas2016shear} and suspending medium \cite{zarraga2000characterization}. Presumably, the resulting alterations to particle surface properties perturb the stress-dependent microphysics and hence $\fAbar$ and $\fBbar$, leading to $\mathcal{O}(1)$ variations in $\eta(\sigma)$.

Our model also predicts a plethora of class 3c flow curves (mapped out fully in the SI \cite{SI}), including a commonly-observed variation of class 3a in which shear thickening is followed by further thinning \cite{Barnes1989,bertrand2002shear,jamali2015microstructure,brown2010generality}. Such behavior arises if constraints on $\mathcal{B}$ are released more slowly than constraints on $\mathcal{A}$ are formed ($\alpha/\beta>1$).

%---------------------------------------- SINGULARITIES ----------------------------------------%
Thus, considering two constraint types gives all three classes of flow curve. In the SI \cite{SI}, we discuss the transition between the classes as $\sigma_{\rm A}/\sigma_{\rm B}$ is varied. In our scheme, system-specific interactions affect the rheology only in how they constrain sliding, rolling and twisting, and how the resulting constraints depend on $\sigma$. Constraints control the jamming point, $\phi_{\rm J}$, Eq.~(\ref{eq:phiJ2}), which is therefore itself $\sigma$ dependent, $\phi_{\rm J}(\sigma)\equiv \sigma_{\rm J}[\fAbar(\sigma),\fBbar(\sigma)]$. At fixed $\phi$, $\eta(\sigma)$ is determined by the ($\sigma$-dependent) distance to jamming $\Delta \phi(\sigma)=\phi_{\rm J}(\sigma)-\phi$, Eq.~(\ref{eq:eta}).

Some interactions do not impose constraints on inter-particle sliding, rolling or twisting, e.g.,~conservative repulsive or attractive interactions (electrostatic, depletion etc.), but still lead to a $\gdot$- or $\sigma$-dependent rheology.
How, then, does one determine whether the rheology is driven by constraints, or by other physics?

To answer this, note that at fixed $\phi$ the system can be in one of two rheological states, depending on $\sigma$. If $\phi_{\rm J}(\sigma)>\phi$, the system flows. If $\phi_{\rm J}(\sigma) \leq \phi$, it jams, i.e., $\eta \rightarrow \infty$. Although non-constraint physics can give rise to such bipartite behavior, our model makes specific predictions for the emergence of viscosity singularities that are not expected to arise generically otherwise. Whenever confirmed, these would rule in constraint physics. 

\begin{figure}[t]
\centering
\begin{minipage}{0.49\columnwidth}
\includegraphics[width=0.86\textwidth]{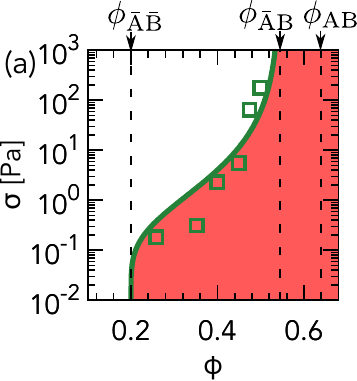}
\end{minipage}
\begin{minipage}{0.49\columnwidth}
\includegraphics[width=\textwidth]{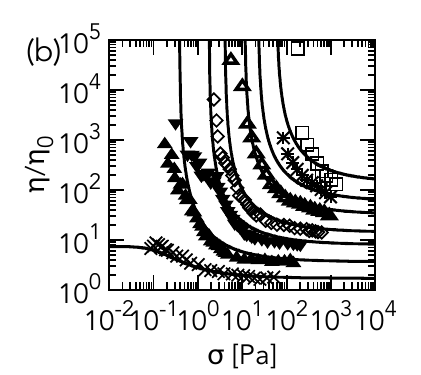}
\end{minipage}
\caption{Singular behavior for class 1. (a) $\sigma$-$\phi$ phase diagram showing jammed ($\phi_{\rm J}\leq\phi$, red) and flowing ($\phi_{\rm J}>\phi$, white) states. Solid curve, jamming boundary $\sigma_{\rm jam}(\phi)$. Vertical dashed lines denote different jamming points, as labelled. Symbols, yield stresses from Ref.~\cite{heymann2002solid}, estimated as the lowest $\sigma$ accessed at each $\phi$. 
(b) Symbols, flow curves from Ref.~\cite{heymann2002solid} at $\phi=0.13$, 0.26, 0.35, 0.40, 0.45, 0.47 and 0.50, from bottom to top; $\eta_0=\SI{0.216}{\pascal.s}$.
Lines, model predictions for $\sigma_{\rm A}\rightarrow 0$, $\sigma_{\rm B}=\SI{1.2}{\pascal}$, $\alpha=1.0$, $\beta=0.5$ and $\phiAB=0.64$, $\phiAbarB=\phiABbar=0.545$ and $\phiAbarBbar=0.20$, at the same $\phi$. 
}
\label{fig:class1}
\end{figure}

We explore these predictions for class 1 and class 3a (see SI \cite{SI} for class 3b) and compare them to literature data. Figure~\ref{fig:class1}(a) shows a typical $\sigma$-$\phi$ ``phase diagram" for class 1 (see caption for parameters). Red states are jammed, $\phi_{\rm J}(\sigma) \leq \phi$, and white states are flowing, $\phi_{\rm J}(\sigma)>\phi$. The boundary of jammed states, $\sigma_{\rm jam}(\phi)$, is defined by $\phi_{\rm J}(\sigma)=\phi$, which we solve for numerically using Eq.~(\ref{eq:phiJ2}-\ref{eq:fBbar}).

Figure~\ref{fig:class1}(b) (lines) shows flow curves at different $\phi$ generated using the same model parameters. The form of $\eta(\sigma)$ (vis-\'a-vis singularities) at any given $\phi$ can be derived by tracing a vertical path with increasing stress in Fig.~\ref{fig:class1}(a). Thus, at some $\phi< \phiAbarBbar$ the system flows at all $\sigma$. At $\phi \geq \phiAbarBbar$, the system is jammed until $\sigma$ exceeds $\sigma_{\rm jam}$, whereupon it flows, i.e., the system has a yield stress $\sigma_{\rm y}=\sigma_{\rm jam}$. This $\sigma_{\rm y}$ increases with $\phi$, diverging at $\phiAbarB<\phiAB$. Thus, we predict jamming at any $\sigma$ if $\phiAbarB \leq \phi < \phiAB$.
 
A yield stress diverging at a $\phi$ substantially below $\phi_{\rm rcp}$ ($\equiv \phiAB$) is evident in many class 1 granular dispersions \cite{heymann2002solid,zhou1995yield}. We show representative data \cite{heymann2002solid} for $R=\SI{2.5}{\micro\metre}$ polymethylmethacrylate (PMMA) spheres in polydimethylsiloxane (PDMS), Fig.~\ref{fig:class1}(b) (symbols). Here, $\sigma_{\rm y}$ [Fig.~\ref{fig:class1}(a) (symbols)] emerges at $\phi \approx 0.2$, which we take as $\phiAbarBbar$, and diverges at $\approx 0.55$, which we take as $\phiAbarB$. Since $\phiAbarB \approx \phi_{\rm rlp}$, $\mathcal{A}={\rm sliding}$ in this system.

\begin{figure}[t]
    \centering
    \begin{minipage}{0.49\columnwidth}
        \includegraphics[width=0.86\textwidth]{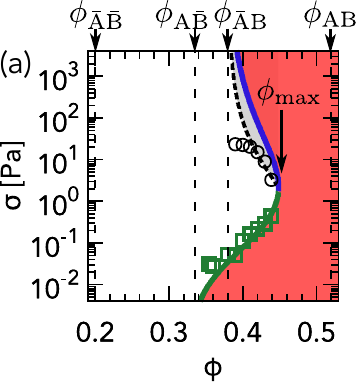}
    \end{minipage}
    \begin{minipage}{0.49\columnwidth}
        \includegraphics[width=1.02\textwidth]{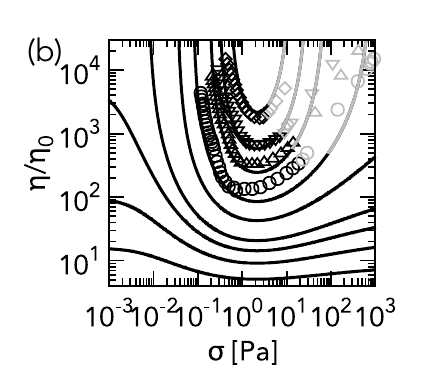}
    \end{minipage}
  \caption{Singular behavior for class 3a. (a) $\sigma$-$\phi$ phase diagram showing jammed (red), stable flowing (white) and unstable flowing (gray) states. Solid line, $\sigma_{\rm jam}(\phi)$ (green=yield stress, blue=re-jamming stress). Dashed curve, boundary of unstable states $\sigma_{\rm uns}(\phi)$. Symbols, data from Ref.~\cite{fall2015macroscopic}: ($\square$) yield stress (estimated as $\sigma$ at the lowest $\gdot$ accessed for each $\phi$) and ($\circ$) the onset of banded flow.
(b) Symbols, flow curves from Ref.~\cite{fall2015macroscopic} at $\phi=0.40$, 0.42, 0.43 and 0.439, from bottom to top; $\eta_0=\SI{1}{\milli \pascal.s}$  \cite{SI}. Gray points correspond to shear-banded states.
Lines, model predictions for $\sigma_{\rm A}=\SI{10}{\pascal}$, $\sigma_{\rm B}=\SI{0.085}{\pascal}$, $\alpha=0.36$, $\beta=0.38$ and $\phiAB=0.52$, $\phiAbarB=0.38$, $\phiABbar=0.335$ and $\phiAbarBbar=0.20$. Volume fractions (from bottom to top), $\phi=0.25$, 0.30, 0.33, 0.35, 0.38, 0.40, 0.42, 0.43, 0.439. Gray parts of the curve are unstable $d\log \eta/d\log \sigma > 1$ states.}
\label{fig:class3a}
\end{figure}

The phase diagram for class 3a is more complex. We plot a representative example for $\phiABbar<\phiAbarB$ in Fig.~\ref{fig:class3a}(a) (see SI \cite{SI} for the qualitatively similar cases of $\phiABbar=\phiAbarB$ and $\phiABbar > \phiAbarB$) and a corresponding set of flow curves at different $\phi$ in Fig.~\ref{fig:class3a}(b) (lines). Note first that there is a $\phi$ window between $\phiAbarB$ and some $\phi_{\rm max}$ in which $\sigma_{\rm jam}(\phi)$, Fig.~\ref{fig:class3a}(a) (solid line), is double-valued. 
For fixed $\phi$ in this window, we predict re-entrant jamming: the system un-jams above a yield stress $\sigma_{\rm y}$ [=lower part of $\sigma_{\rm jam}(\phi)$, green] and subsequently re-jams at a higher stress [=upper part of $\sigma_{\rm jam}(\phi)$, blue] \footnote{We compare the different physics behind Fig.~\ref{fig:class1}(a) and \ref{fig:class3a}(a) in the SI \cite{SI}.}.
Secondly, there is a critical $\phi_{\rm max}<\phiAB$ above which the system is jammed at all $\sigma$. Note that $\phi_{\rm max}$ is {\it not} associated with any divergence; rather this is the $\phi$ at which the yield stress (green) and re-jamming stress (blue) become equal. Thus, as $\phi \to \phi_{\rm max}^-$, the stress window of flowing states shrinks, Fig.~\ref{fig:class3a}(b), and vanishes at $\phi_{\rm max}$.

There is a region in Fig.~\ref{fig:class3a}(a) (gray) in which $d\sigma/d\gdot<0$, corresponding to $d\log \eta/d\log \sigma>1$ in Fig.~\ref{fig:class3a}(b) (gray lines). Here, shear flow is unstable \cite{yerushalmi1970stability}. The boundary of this region, $\sigma_{\rm uns}(\phi)$ (black dashes), corresponds to solutions of $d\log \eta/d\log \sigma=1$ and meets the yield (green) and re-jamming (blue) stresses at $\phi_{\rm max}$. Under imposed $\gdot$, $\sigma_{\rm uns}$ is the maximum stress for stable flow.

Most systematic work on class 3a singular behavior has been done for aqueous cornstarch dispersions ($R \approx \SI{7}{\micro\metre}$) \cite{fall2015macroscopic,madraki2017enhancing,hermes2016unsteady,peters2016direct}. Under imposed $\sigma$, the only study of the entire phase diagram \cite{peters2016direct} found pronounced shear thinning followed by a completely rigid state, consistent with our prediction. Typical phenomenology under imposed $\gdot$ is shown in Fig.~\ref{fig:class3a}(b) (symbols)~\cite{fall2015macroscopic}. Cornstarch flows steadily above a yield stress [$\square$ in Fig.~\ref{fig:class3a}(a)] until some higher $\gdot$ limit is reached, above which the suspension separates into shear bands, Fig.~\ref{fig:class3a}(b) (gray symbols). The stress at this upper limit plausibly corresponds to our $\sigma_{\rm uns}(\phi)$ line. Indeed, the experimental $\sigma_{\rm uns}$ [$\circ$ in Fig.~\ref{fig:class3a}(a)] and $\sigma_{\rm y}$ ($\square$) appear to converge at $\phi_{\rm max} \approx 0.45 \ll \phi_{\rm rcp}$. As we would predict, flow was found impossible for $\phi$ beyond this limit. The existence of a continuum of fully jammed states at $\phi_{\rm max} < \phi < \phi_{\rm rcp}$, a surprise for the authors of Ref.~\cite{fall2015macroscopic}, emerges naturally from our model.

Note that all classes of granular suspension flow curve can be obtained by appealing to a single constraint, but with $f(\sigma)$ mimicking $\eta(\sigma)$ in each case, e.g., a peaked $f(\sigma)$ will give class 3b flow curves \footnote{For sliding, lubricating films rupturing above a critical stress to form load-dependent frictional contacts can give a peaked $f(\sigma)$~\cite{chatte2018shear}.}. This contrasts with our appeal to two constraints with simple, generic forms for $f$, Eqs.~(\ref{eq:fAbar}) and (\ref{eq:fBbar}), to give all classes of flow curves. A single constraint can lead to phase diagrams that are topologically similar to Fig.~\ref{fig:class1}(a) and Fig.~\ref{fig:class3a}(a); however, singular behaviour can arise only in a relatively narrow $\phi$ window, $\phi_{\bar{\rm A}} \leq \phi < \phi_{\rm A}= \phi_{\rm rcp}$. 

We should emphasize that our framework has been constructed for {\it granular} suspensions. Recently, class 3a flow curves, Fig.~\ref{fig:class3a}(b), have been observed in simulations of {\it Brownian} particles with inter-particle friction (=~constrained sliding) and conservative potential attraction \cite{pednekar2017simulation}, giving a yield stress that depends much more weakly on $\phi$ (roughly $\propto \phi^2$) than is observed in granular systems, Fig.~\ref{fig:class3a}(a) ($\square$). If the potential attraction is strong enough, the yield stress ``masks" shear thickening and the system becomes class 1, Fig.~\ref{fig:class1}(b). However, the physics of yielding in a system where constraints act in the presence of Brownian motion is little understood and may be quite distinct from that explored here.

Before concluding, we turn briefly to microphysics. While much is known about the tribology of dry grains, little is known about the effect of a solvent on sliding, rolling and twisting resistance. If, like others \cite{wyart2014discontinuous,boyer2011unifying}, we assume that `dry tribology' is possible in a suspension, then one  realization of our scheme is the finite-area, adhesive contact with asymmetric pressure distribution between frictional particles \cite{dominik1995resistance, heim1999adhesion}. Such a contact is pinned by surface roughness, leading to a critical torque $M^*$ for ``peeling" particles apart and $\sigma_{\rm B} \sim M^*/R^3$. 
Interestingly, simulations of adhesive, frictional dry grains \cite{liu2017equation} are consistent with the observation of $\phiAbarB \approx \phiABbar$ for cornstarch in Fig.~\ref{fig:class3a}(a). On the other hand, our A\Bbar~state (sliding unconstrained, rolling constrained) has no obvious dry-granular analogue, suggesting that the contact physics of dispersed particles holds surprises.

To conclude, we have presented a constraint-based phenomenological model for granular dispersion flow that predicts all known classes of experimental flow curve. Several non-trivial predictions for the emergence of singular behavior follow, which are borne out by literature data. Our notion that system-specific microphysics enters the rheology only on the level of constraints is a powerful one. It allows disparate systems to be captured in a single treatment agnostic of microphysics, much like the way multifarious details are subsumed into a ``tube constraint'' in polymer rheology \cite{doi1988theory}. Many  challenges remain, including making a formal link between constraints, jamming and dissipation \cite{Lerner2012,trulsson2017effect} and probing the microphysics of constraints in the presence of solvent.
 
{\bf Acknowledgements} BMG was funded by the EPSRC (EP/J007404/1). JAR was funded by EPSRC SOFI CDT (EP/L015536/1) and AkzoNobel. DJMH and EB were funded by Mars Chocolate UK Ltd. WCKP was funded by the EPSRC (EP/J007404/1 and EP/N025318/1). We thank John Brady, Mike Cates, Annie Colin, Guillaume Ovarlez and Eric Cl\'ement for discussions.

{\bf Author contributions} BMG, JAR, EB and DJMH performed experiments motivating the model. JAR and WCKP conceptualized the work. JAR and BMG performed calculations. BMG, JAR and WCKP produced the final manuscript.

\bibliographystyle{apsrev4-1}
\bibliography{constraints_arXiv}

\end{document}